\documentclass[a4paper,12pt]{article}
\usepackage[utf8]{inputenc}
\usepackage[a4paper, margin=2.5cm]{geometry}
\usepackage{booktabs}
\usepackage{graphicx}
\usepackage{hyperref}
\usepackage{listings}
\usepackage{amsmath}
\usepackage{verbatim}
\usepackage{color}
\usepackage{wrapfig}
\usepackage{lineno}
\usepackage{array} 
\usepackage{subcaption}
\usepackage{caption}
\usepackage{listings}
\usepackage{xcolor}
\usepackage{colortbl}
\usepackage{hyperref}
\usepackage{authblk}
\usepackage{longtable}
\usepackage[style=nature,url=true,doi=true,maxbibnames=3, defernumbers=true]{biblatex}
\bibliography{bibliography.bib}
\usepackage{setspace}
\usepackage{pdfpages}

\author[1,2,+]{\normalsize Ramit Debnath}
\author[1*]{\normalsize Pengyu Zhang}
\author[1*]{\normalsize Tianzhu Qin}
\author[2]{\normalsize R. Michael Alvarez}
\author[1]{\normalsize Shaun D. Fitzgerald}
\title{\vspace{-2.5 cm} \Large Deciphering public attention to geoengineering and climate issues using machine learning and dynamic analysis}

\affil[1]{University of Cambridge, Cambridge, CB2 1PZ, UK}
\affil[2]{California Institute of Technology, Pasadena, 91125, US}
\affil[*]{Equal contribution}
\affil[+]{Corresponding author: rd545@cam.ac.uk}

\date{}
\begin{document}

\maketitle

\begin{abstract}

As the conversation around using geoengineering to combat climate change intensifies, it is imperative to engage the public and deeply understand their perspectives on geoengineering research, development, and potential deployment. Through a comprehensive data-driven investigation, this paper explores the types of news that captivate public interest in geoengineering. We delved into 30,773 English-language news articles from the BBC and the New York Times, combined with Google Trends data spanning 2018 to 2022, to explore how public interest in geoengineering fluctuates in response to news coverage of broader climate issues. Using BERT-based topic modeling, sentiment analysis, and time-series regression models, we found that positive sentiment in energy-related news serves as a good predictor of heightened public interest in geoengineering, a trend that persists over time. Our findings suggest that public engagement with geoengineering and climate action is not uniform, with some topics being more potent in shaping interest over time, such as climate news related to energy, disasters, and politics. Understanding these patterns is crucial for scientists, policymakers, and educators aiming to craft effective strategies for engaging with the public and fostering dialogue around emerging climate technologies.

\end{abstract}

\newpage

\section{Introduction}

Geoengineering, climate engineering, or climate intervention refers to a set of emerging technologies that could impact the environment and partially offset some of the impacts of climate change \cite{Keith_2021}. Most geoengineering techniques can be grouped into two categories: solar radiation management (SRM) or solar geoengineering (SG), and greenhouse gas removal (GGR) or carbon geoengineering \cite{Shepherd_2012}. There are some exceptions such as the erection of seabed curtains to reduce the rate of supply of deep warm saline water to the grounding line of glaciers \cite{lockley2020glacier}. SRM techniques aim to reflect a small proportion of the sun’s energy back into space, counteracting the temperature rise caused by increased levels of greenhouse gases (GHG) in the atmosphere, which absorb energy and raise temperatures. GGR techniques aim to remove carbon dioxide or other GHGs from the atmosphere, directly countering the increased GHG effect and ocean acidification \cite{gloenvcha2023, Sovacool_2021}. With the global urgency to limit GHG emissions, these approaches are receiving increased, at times controversial, attention from scientists, policymakers, investors, and the general public \cite{gloenvcha2023, Debnath_Fitzgerald2023, Keith_2021, Schellnhuber_2011, Anderson_Peters_2016, Sovacool_2021, DebnathEbanksMohaddes2023}.

In the current context, some types of GGR have been employed with mixed results, and SRM remains a theoretical approach, making them currently uncertain and risky for deployment at scale. In addition, scientists and decision-makers have up until now largely driven discussions about GGR and SRM. Recent studies show that it is crucial to engage with the public, engage citizens in deliberations, and establish communication pathways to map interests, attitudes, and concerns about these emerging technologies \cite{Buck_2016, Bellamy_2018, Colvin_Kemp_Talberg_De2019}. This kind of multilateral cooperation could help stop or reduce any problems that might come up when SRM and GGR technologies are used on a large scale \cite{Colvin_Kemp_Talberg_De2019, Forster_Vaughan_Gough2020, gloenvcha2023}. However, the effective measurement of public attitudes and opinion dynamics across SRM and GGR is challenging due to three issues. First, \textit{the perception-knowledge gap}: surveys have found that people know very little about GGR and even less about SRM, which shapes a wide-ranging perception from support to entrenched opposition \cite{Cox_Spence_Pidgeon_2020, Pianta_Rinscheid_Weber_2021, Pidgeon_Spence_2017, Sweet_Schuldt2021, Whitmarsh_Xenias_Jones_2019, Sovacool_2021}. Second, \textit{data scarcity}: traditional survey methods can get tremendously resource-intensive when scaling up to capture public attitudes, attention and perceptions of climate action \cite{DebnathSovacool2022, Debnath_van2023}. Third, \textit{measurement}: using surveys for studying attitudes and opinions about SRM and GGR is challenging because there is little existing research on how to appropriately frame and word questions about these topics.

Recent advances in computational social science help address part of the above challenge by leveraging unstructured text-based data in social media platforms, newspaper archives, and opinion-sharing websites like Reddit, Quora, etc. \cite{Debnath_van2023, DebnathSovacool2022, LazerPentlandFowlerGutmann2009}. In this study, we conducted a similar unstructured, data-driven investigation of interests and attitudes towards geoengineering. Here, we used a time-series dataset from 2018 to 2022 constructed using Google Trends, online BBC News, and the New York Times (NYTimes) articles (combined articles = 30773) covering topics associated with climate change and geoengineering to test whether attitudes towards climate issues covered in news and geoengineering are linked and whether they influence sentiments towards SRM and GGR.

We employed a data-driven machine learning (ML) and time-series analysis of this news dataset of climate issues to gauge public interest in climate action and related technological developments \cite{Happer_Philo_2015, JunshengBanna_2019}. This methodological choice is predicated on the assumption that media representations both mirror and influence societal concerns and priorities, thereby acting as an essential barometer for gauging public interest \cite{Happer_Philo_2015, JunshengBanna_2019, Schafer_Painter_2020, BergquistJagers2022, ChinnSoroka_2020}. The first phase of our analysis seeks to quantitatively determine the relationship between the volume of media coverage concerning various environmental and climate issues, and the level of public attention and engagement with geoengineering concepts. We use the five years of publicly available Google Trends search data to examine the above relationship using time-series regression analysis (\cite{MavraganiOchoaTsagarakis2018, Durmulu_2017, Jun_Yoo_Choi_2018, Choi_Varian_2012}). We reference this as Hypothesis 1 (H1):
\begin{itemize}
    \item H1: Coverage of climate and environmental issues is correlated with public interest in geoengineering-related content.
\end{itemize}

The second phase of our analysis unpacks the topics embedded in the news coverage on the BBC and the NYTimes, identifying specific environmental issue topics that have a more frequent representation in the media discourse. We use machine learning (ML), natural language processing (NLP)-led topic modelling (like BERTtopic), and statistical correlation analysis to sort and rate the importance of each topic in media coverage. Due to the temporal nature of our data, we also employ time-series regressions to study the dynamic trends in media coverage and public interest. This helps us find the topics that raise the most public awareness in geoengineering and may inform the design of appropriate public engagement strategies. This is Hypothesis 2 (H2): 
\begin{itemize}
    \item H2: Different environmental and climate challenges prompt different levels of public attention to geoengineering.
\end{itemize}

Finally, we delve into the association/correlation of sentiment expressed in news coverage of climate issues with public perceptions of geoengineering. The hypothesis is that the emotional tone of media reporting — whether optimistic, pessimistic, neutral, or alarmist — is associated with shaping public attitudes towards geoengineering over time. By examining the dynamics of sentiment on the news coverage of climate issue  across a 5 year period, we evaluate the level of public attention and acceptance of geoengineering as a potential viable strategy for climate action. In this context, we test Hypothesis 3 (H3): 

\begin{itemize}
    \item H3: Differential sentiment tones employed across various environmental and climate news stories are associated with varying levels of public attention and attitudes towards geoengineering. This is reflected across the time period of the dataset.

\end{itemize}

In summary, through this data-driven and multi-method approach, this paper endeavours to contribute to a deeper understanding of the complex interplay between media representation of environmental issues and public engagement with, and attitudes towards geoengineering, fulfilling an important knowledge and methodological gap between public perception, risk management, and science communication of futuristic GGR and SRM technologies.

\section{Data and Methods}

This paper uses a data-driven methodology to examine the hypotheses (H1 to H3) using a combination of natural language processing (NLP), machine learning (ML), and time-series statistical analysis. For NLP, standard pre-processing steps were followed as per the best practice guidelines that consist of tokenization, stemming, and lemmatization (\cite{DebnathSovacool2022, Debnath_Fitzgerald2023}). Tokenization refers to breaking down the given text into smaller units in a sentence called tokens. Stemming in NLP is a morphological technique that breaks words into their root form. Lemmatization is a technique used to reduce inflectional forms of words to a common base form. In this NLP pre-processing pipeline, stopwords were removed. Stopwords are the most common words in any language (like articles, prepositions, pronouns, conjunctions, etc.), which do not add much information to the text.

Following this NLP pre-processing procedure, we performed ML-led (deep learning) topic modelling using BERTopic \cite{grootendorst2022bertopic}, which leverages transformers and c-TF-IDF to create dense clusters of associated topics from text corpus, allowing for easily interpretable topics while keeping contextually important words in the topic descriptions \cite{grootendorst2022bertopic}. Next, using a similar approach, we estimated the sentiment scores of the text documents using BERT sentiment analysis.

Finally, we used statistical techniques such as linear regression and time-series regression analysis using autoregressive (AR), ARIMA, and SARIMA models to investigate the dynamic relationship between media coverage of climate and environmental topics and public interests and attitudes towards geoengineering. Further details are presented below.

\subsection{Data}

\subsubsection{Google Trends} 

Google Trends data is a sample of Google search data, including top search queries, across various regions and languages. Google Trends data are anonymized, categorised by topic, and aggregated. This allows users to measure interest in a particular topic across searches, from around the globe right down to city-level geography \cite{nuti2014use, choi2012predicting, jun2018ten}.

To extract data from Google Trends, users first identify and input various search terms of interest, followed by setting the desired time range for analysis. The Google Trends Index then provides data on the search frequencies for the specified term within Google over the time period as a csv file. The Google Trends Index involves an analysis by Google Trends of a set of Google web searches. This process aims to quantify the volume of searches conducted within a specified time frame. The results are then normalised, with values ranging from 0 to 100 \cite{GoogleTrend}.

In this paper, we use Google Trends data from 2018–2022 to examine public interests and attitudes towards geoengineering and climate issues (as illustrated in SI Figure 1). We extract key terms used in this context based on \cite{gloenvcha2023}, further illustrated in Table \ref{tab:tab1}. Google specifies that its trend data is not a scientific poll and should not be confused with polling data as it is a reflection of search interest in particular topics. However, recent studies have shown a growing interest in its use to study public attitudes and in econometric analysis \cite{MavraganiOchoaTsagarakis2018}. For example, Polden, Robinson, and Jones (2023) have used Google Trends to assess public awareness of food labelling policies in the UK \cite{Polden_Robinson_Jones_2023}. Ryu and Min (2020) have used trend data to measure public perceptions of air quality by developing a public perception model using internet search volume \cite{Ryu_Min_2020}. The use of Google Trends data in public perception research is an emerging topic, and our paper contributes significantly to this timely discussion.

\begin{table}[htbp]
    \centering
\caption{Key terms used in the study to construct the text corpus from the BBC and the NYTimes online news articles that includes news stories, comments, opinion and editorial notes ($n$ = 30773) between 2018 and 2022 [Note: Some of the terms are not mutually exclusive.]}
\label{tab:tab1}
    \begin{tabular}{|>{\centering\arraybackslash}p{0.3\linewidth}|>{\raggedright\arraybackslash}p{0.7\linewidth}|} \hline 
         Terms& Simple definition\\ \hline 
         Geoengineering& This is a broad term encompassing large-scale interventions in the Earth's climate system to counteract climate change. Its inclusion is crucial as it's a comprehensive term covering various climate intervention technologies. \cite{keith2000geoengineering, caldeira2013science}\\ \hline 
         Afforestation& This refers to the planting of trees in an area where there was no previous tree cover.  It's a natural form of carbon capture, relevant for its potential to mitigate climate change. \cite{anderson1987economics, arora2011small}\\ \hline 
         Biochar& This is a form of charcoal produced from biomass and used as a soil amendment. It's significant for its ability to sequester carbon and improve soil health, thus contributing to climate mitigation.\cite{weber2018properties}\\ \hline 
         Bioenergy& This involves producing energy (electricity, heat, or fuel) from biomass. It's included for its role in renewable energy discussions and its potential to reduce greenhouse gas emissions.\cite{slade2014global, bioenergy2009bioenergy}\\ \hline 
         Carbon capture and storage (CCS)& CCS technology captures carbon dioxide emissions from sources such as power plants and stores it underground.  It's a key term in discussions about reducing industrial carbon emissions.\cite{boot2014carbon}\\ \hline 
         Direct Air Capture (DAC)& DAC involves technologies that capture carbon dioxide directly from the ambient air.\cite{zhu2022recent, kumar2015direct}\\ \hline 
         Marine Cloud Brightening& This concept involves spraying seawater and allowing natural convection to carry additional salt crystals into marine clouds, to increase their reflectivity, thus reducing solar radiation and cooling the Earth. It is included for its potential as a solar radiation management technique.\cite{latham2012marine}\\ \hline 
         Solar Geoengineering (SG)& SG refers to technologies designed to reflect a small portion of solar radiation back into space.\cite{kreith1978principles, goswami2022principles}\\ \hline 
 Stratospheric Aerosol Injection& This involves injecting reflective particles into the stratosphere to reflect sunlight and cool the planet. It is a controversial yet prominent idea in solar radiation management discussions.\cite{hulme2012climate}\\ \hline
 Ocean Iron Fertilization&Iron fertilization is the deliberate addition of iron-containing substances (such as iron sulphate) to water surfaces that don't have enough iron to encourage the growth of phytoplankton. This is meant to boost biological productivity and speed up the removal of carbon dioxide (CO2) from the air. \cite{latham2012marine}\\\hline
 Ocean Alkanility Enhancement&Ocean alkalinization is an approach to carbon removal by adding alkaline substances to seawater to help the ocean's natural carbon sink work better. Minerals such as olivine or man-made substances such as lime or some industrial waste could be among these substances. \cite{EisamanSTP_2023}\\\hline
    \end{tabular}

\end{table}

\subsubsection{News Corpus}

In this paper, we only used publicly available APIs for English-language news data extraction. The New York Times (NYTimes) \cite{von2018sourcing, kiousis2004explicating} and the British Broadcasting Corporation (BBC) news \cite{tian2005framing, hossain2022computer, gomez1994relevance} offer such APIs for academic use and are our primary data sources. 

The BBC and the NYTimes are two of the most visited news websites in the world with BBC having over 1.1 billion visits alone in 2023 \cite{news2024}. They cover the latest news across the socio-political, environmental, and technological spectrum and have a diverse readership. Therefore, it served as an appropriate dataset for our analysis.  

\begin{figure*}[htbp]
\centering
\includegraphics[width=0.9\linewidth]{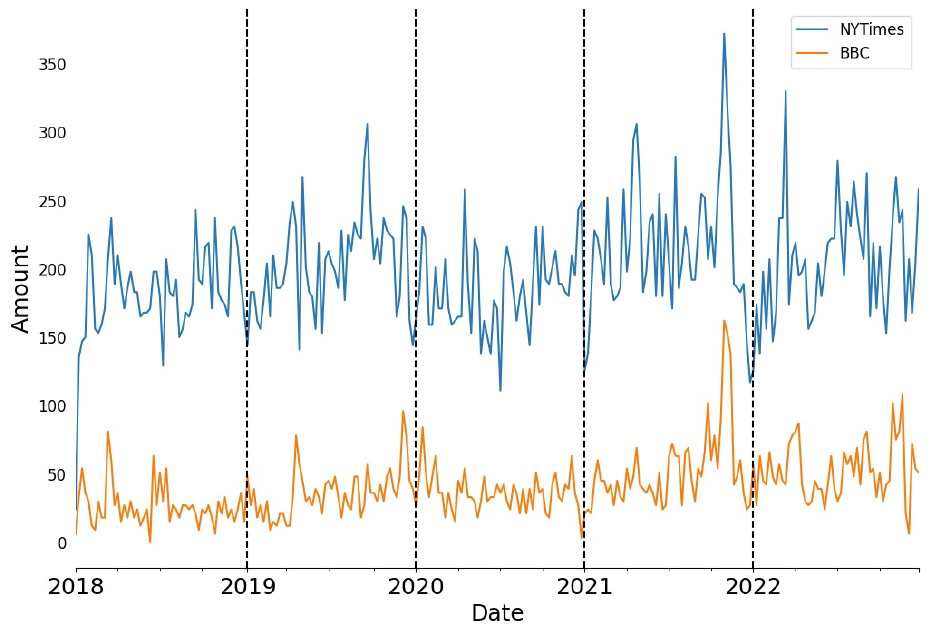}
\caption{\textbf{The BBC and the NYTimes news temporal trends for 2018–2022.} This figure shows the temporal trends of news articles produced per week containing terms associated with geoengineering and climate change, as per Table 1. The total number of relevant news articles for our analysis is 30,773.} \label{fig:1}
\end{figure*}

Figure \ref{fig:1} illustrates a weekly temporal trend between 2018 and 2022 for both the BBC and the NYTimes articles. After manual filtration of relevant news items based on keywords in Table \ref{tab:tab1}, we arrived at 5333 BBC articles and 25440 NYTimes articles with geoengineering and climate-related content. It includes news stories, comments, opinion and editorial notes (total $n$ = 30,773, see Supplementary Information (SI) Table 1 for descriptive statistics). We use this cumulative $n$ as our data corpus for the NLP analysis, described in section 2.2.

\subsection{Methods}

\subsubsection{Overview}

Our model consists of two primary data-driven components: a Google Trends analysis and a media NLP and ML-led framing analysis. These components work in tandem to investigate the relationship between media coverage and public interest in geoengineering, climate and environmental issues (collected using the keywords in Table \ref{tab:tab1}).

In order to analyse the large collection of media articles related to geoengineering and environmental issues, we employ NLP techniques that provide in-context learning, specifically BERT topic modelling \cite{abuzayed2021bert, glazkova2021identifying, peinelt2020tbert, thompson2020topic} and BERT sentiment analysis \cite{hoang2019aspect, xu2019bert, sousa2019bert, alaparthi2021bert}. BERT topic modelling identifies the dominant topics discussed in the articles, capturing the key themes and concepts surrounding geoengineering. Furthermore, BERT sentiment analysis gauges the emotional tone and opinions expressed in the articles, providing insights into the embedded sentiment in news articles towards geoengineering and environmental issues (further explained below).

The approach underscores a bidirectional analysis where the relationship between media portrayal and public awareness is quantified, allowing for an evaluation of how media influences public discourse and sentiment on critical climate and environmental issue topics. This methodological synergy between content analysis and trend assessment aims to offer a robust understanding of the media's role in shaping public engagement with geoengineering.

\subsubsection{Google Trends Correlation}

We identified  the terms which were most correlated with `geoengineering' in Google Trends searches between 2018 and 2022. Here, we used the Pearson Correlation (PC) coefficient to measure the strength of the association of searched terms associated with geoengineering on Google \cite{rangkuti2018sentiment, vallejo2022evaluating}.

\subsubsection{Word2Vec Distance}

For semantic analysis, we utilise the concept of distance in Word2Vec space, constructed using news media data, to measure the relationship between geoengineering and various climate issues. Word2Vec is a ML model that transforms words into multi-dimensional vectors, capturing their contextual meanings based on their usage in large text corpora \cite{mikolov2013efficient, ma2015using}. By analysing these vectors in a constructed Word2Vec space, we can quantitatively assess the 'distance' between the terms 'geoengineering' and specific environmental issues. For example, Krishnan and Anoop (2023) constructed ClimateNLP using Word2Vec to study public sentiments towards climate change \cite{Krishnan_Anoop_2023}. Stede et. al. (2023) used this ML model to extract the framing of climate change in Nature and Science journal editorials \cite{StedeCSS2023}. 

The Word2Vec model employs neural networks to convert words from our news media dataset into multi-dimensional vectors. Mathematically, each word \( w \) is represented as a high-dimensional vector \( \vec{v}_w \) in a continuous vector space. The position of each word in this space is determined based on its contextual relationships with other words in the dataset.

To assess the relationship between 'geoengineering' and various climate issues, we calculate the Euclidean distance between their corresponding vectors. Its empirical form is illustrated in the Supplementary Information (SI). 



This `distance' is not just a measure of physical separation in vector space but rather represents the degree of semantic and contextual association between these terms. A shorter distance indicates a stronger relationship or higher relevance between geoengineering and a given environmental issue, as reflected in media discourse. Conversely, a greater distance suggests a weaker or less frequent association \cite{mikolov2013efficient, ma2015using}.






\subsubsection{BERT Topic}

Google created the BERT (Bidirectional Encoder Representations from Transformers) model, which is one approach to NLP using deep learning \cite{devlin2018bert}. It revolutionises how machines interpret human language by focusing on the context of words in a sentence. Unlike previous models that analysed words in sequence, BERT interprets words based on their surrounding words, providing a deeper understanding of language nuances \cite{jawahar2019does}. This makes BERT highly effective for tasks like sentiment analysis, question answering, and language translation. Its bidirectional nature allows it to grasp the full context of a sentence, leading to more accurate interpretations and predictions in various language processing applications. In this paper we have used \emph{bertopic} v0.16.0 \cite{grootendorst2022bertopic}.

The BERTopic model leverages BERT's transformer architecture for creating document embeddings that are contextually rich. This method allows for capturing the nuanced meanings and relationships in text, which are crucial for accurate topic modeling. The model then uses dimensionality reduction on these embeddings to efficiently cluster, which means it groups documents that are similar \cite{abuzayed2021bert, glazkova2021identifying, peinelt2020tbert}. By applying class-based TF-IDF, BERTopic can extract distinct topics from each cluster, providing a more detailed and representative understanding of the text corpus \cite{grootendorst2022bertopic}. This approach enables BERTopic to dynamically adapt to varying contexts and datasets, making it highly effective for extracting and understanding complex topics in large text corpora \cite{grootendorst2022bertopic}.

\subsubsection{BERT Sentiment}

Transformer-based pre-trained models like BERT are especially helpful in extracting meaningful topics from text data as they contain more accurate representations of words and sentences. By fine-tuning BERT on sentiment-labelled data, our model learns to classify text as positive, negative, or neutral. This involves feeding text data into the pre-trained BERT model and training it further on a dataset specific to sentiment analysis \cite{hoang2019aspect, xu2019bert, sousa2019bert, alaparthi2021bert}. In this paper, we use the bert-base-multilingual-uncased model from Hugging Face \cite{Hugginface_2020}. 

This BERTsentiment model is fine-tuned for sentiment analysis on product reviews in six languages: English, Dutch, German, French, Spanish, and Italian \cite{Hugginface_2020}. This model is intended for direct use as a sentiment analysis model for text analysis with accuracy for English language $\sim$67\%. The output is a sentiment prediction, reflecting the context-specific positive, negative or neutral tone of the input text. The result for each sentence is a value within 1-5, and the result for each news text is the average sentiment score for all the sentences within the news. This approach leverages BERT's in-context learning capabilities for effective sentiment analysis \cite{hoang2019aspect, xu2019bert, sousa2019bert, alaparthi2021bert}.

\subsubsection{Regression and Interactions}

In our study, we employ linear regression analysis \cite{stock2020introduction} to investigate the relationship between media coverage of environmental and climate topics by the BBC and NYTimes and public interests in geoengineering approaches shown in the Google Trends data. This statistical method enables us to quantify the strength and direction of this relationship.

We used BERTopics and BERTsentiment scores as the independent variables, with Google Trends Index scores on topics associated with `geoengineering' as the dependent variables. One assumption is that the Google Trend Index is a proxy for public interest in the topic, with weights ranging from 0 (least interest) to 100 (highest interest).

We ran two separate regression models for results from both BERTopics and BERTsentiment, namely, Model 1 and Model 2; empirical detail are illustrated in the SI.











\subsection{Time-series regressions and dynamic relationships}

To further examine whether the news articles from the BBC and NYTimes lead to higher rates of Google searches related to geoengineering over time, we used three time-series regression models: autoregressive (AR), autoregressive integrated moving average (ARIMA), and seasonal autoregressive integrated moving average (SARIMA). The choice of using these models is motivated by previous research using Google Trends data for measuring public interests in conservation and ecosystems \cite{nghiem2016analysis}, health awareness \cite{hao2019evaluating}, and various topics in epidemiology, stock market, and tourism, for example. \cite{mavragani2018assessing}.

In all regression models conducted in this section, the dependent variable was the Google Trends Index, and the explanatory variables were the significant lags in the number of English news articles across the significantly influential topics identified in Section 2.2. Model specifications are presented in the SI.

\section{Results}

\subsection{Environmental topics and geoengineering interests}

\begin{figure*}[htbp]
\centering
\includegraphics[width=1.0\linewidth]{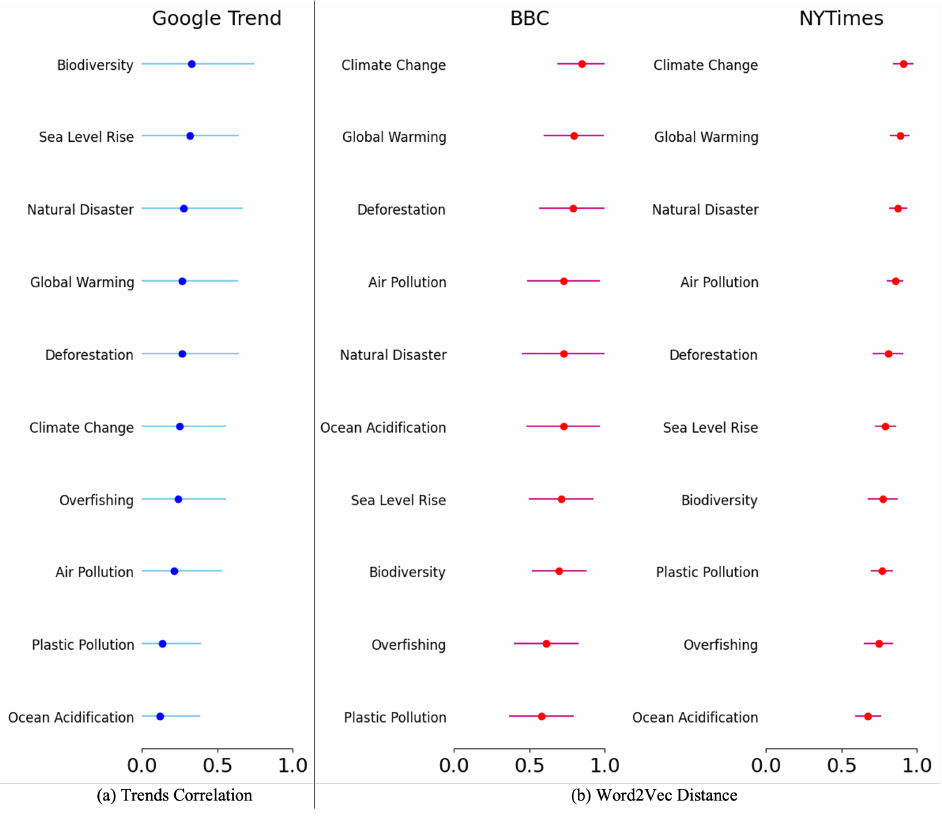}
\caption{ \textbf{Relationship between attention to environmental issues and geoengineering for both the public and media.} (a) The correlation between public interest in various climate topics (represented on the y-axis) and public interest in geoengineering is illustrated using data sourced from Google Trends. The x-axis values represent weighted Pearson correlation coefficients (R$^{2}$), with the dots and lines illustrating the mean and 95\% confidence level. (b) Correlations between media coverage on different environmental issues in the BBC and NYTimes (y-axis) and on geoengineering are calculated as the Word2Vec distance between words. We report the Eucledian distance at 95\% confidence levels.}\label{fig:2}
\end{figure*}

Figure \ref{fig:2} illustrates the relationship between the public's online engagement with various environmental challenges and geoengineering, and the media's coverage of these topics. Using the Google Trends Index data, we find greater interest (R$^{2}\geq$0.5, 95\% CI) across topics like `biodiversity',`sea level rise', `natural disaster', `global warming', `deforestation', and `geoengineering'. Other topics associated with geoengineering interests are climate change, overfishing, air pollution, plastic pollution, and ocean acidification, which also show a R$^{2}$ (95\% CI) value between 0.1 and 0.4 (see Figure \ref{fig:2}a). The strength of the correlation indicates a high level of public interest and engagement with these topics on Google Trends.

Similarly, using the Word2Vec algorithm and Euclidean distance estimation (at 95\% CI), we show the association of topics on environmental issues and geoengineering in the BBC News and NYTimes data corpus. Smaller distances suggest a tighter conceptual linkage, as seen with terms like `climate change' and `global warming', which are closely associated with geoengineering in both the BBC and the New York Times news content (see Figure \ref{fig:2}b). This semantic closeness suggests that as the media discusses these environmental challenges, there can be an implicit or explicit connection to various contexts around geoengineering.

For example, semantically, geoengineering is close to climate change and global warming in the BBC and NYTimes news corpus, indicating that they are likely (at 95\% CI) to appear in the same context as plastic pollution or ocean acidification (see Figure \ref{fig:2}b). Therefore, we observe a certain degree of semantic overlapping between Google Trends searches and media coverage of geoengineering and environmental issues, which supports our \textit{hypothesis 1 (H1)} with the assumption that the Google Trends Index can be a proxy for public interests on that topic. We further report the time-based effect on public attention to geoengineering in Section 3.5, that further strengthens H1.

\subsection{Topic overlap in environmental issues and geoengineering}

BERT-based topic modelling analysis of the 30,773 news articles resulted in 266 topic clusters. Dimensional reduction using hierarchical  clustering followed by a manual selection resulted in 13 key topic clusters: geoengineering, art, disaster, economy, education, energy, medical, nature, politics, pollution, religion, society, and technology. These are illustrated in Table \ref{tab:tab2}.

\begin{table}[htbp]
\centering
\caption{Sub-topics from the BERT model are classified into following topics.}
\label{tab:tab2}
\begin{tabular}{|c|cc|}
\hline
Manually Classified Topic & \multicolumn{2}{c|}{Sub-Topics from BERT topic model}                                                                                                                                                                                                         \\ \hline
geoengineering            & \multicolumn{1}{c|}{\cellcolor[HTML]{FFFFFF}\begin{tabular}[c]{@{}c@{}}carbon\_co\_\\ dioxide\_removal\end{tabular}}        &  \\ \hline
disaster                  & \multicolumn{1}{c|}{\cellcolor[HTML]{FFFFFF}\begin{tabular}[c]{@{}c@{}}fire\_fires\_\\ california\_wildfires\end{tabular}}  & \cellcolor[HTML]{FFFFFF}\begin{tabular}[c]{@{}c@{}}amazon\_deforestation\\ \_brazil\_indigenous\end{tabular}     \\ \hline

energy                    & \multicolumn{1}{c|}{\cellcolor[HTML]{FFFFFF}oil\_gas\_prices\_energy}                                                               & \cellcolor[HTML]{FFFFFF}\begin{tabular}[c]{@{}c@{}}electric\_cars\_\\ vehicles\_car\end{tabular}                \\ \hline
nature                    & \multicolumn{1}{c|}{\cellcolor[HTML]{FFFFFF}\begin{tabular}[c]{@{}c@{}}ice\_glacier\_\\ glaciers\_greenland\end{tabular}}           & \cellcolor[HTML]{FFFFFF}trees\_tree\_forest\_forests                                                            \\ \hline
politics                  & \multicolumn{1}{c|}{\cellcolor[HTML]{FFFFFF}\begin{tabular}[c]{@{}c@{}}glasgow\_cop26\_\\ country\_climate\end{tabular}}       & \cellcolor[HTML]{FFFFFF}\begin{tabular}[c]{@{}c@{}}border\_migrants\_\\ immigration\_immigrants\end{tabular}     \\ \hline
pollution                 & \multicolumn{1}{c|}{\cellcolor[HTML]{FFFFFF}\begin{tabular}[c]{@{}c@{}}plastic\_bags\_\\ recycling\_waste\end{tabular}}             & \cellcolor[HTML]{FFFFFF}air\_pollution\_quality\_study                                                          \\ \hline
society                   & \multicolumn{1}{c|}{\cellcolor[HTML]{FFFFFF}\begin{tabular}[c]{@{}c@{}}meat\_food\_\\ eat\_plantbased\end{tabular}}                  & \cellcolor[HTML]{FFFFFF}\begin{tabular}[c]{@{}c@{}}facebook\_twitter\_\\ data\_ads\end{tabular}                 \\ \hline
medical                   & \multicolumn{1}{c|}{\cellcolor[HTML]{FFFFFF}\begin{tabular}[c]{@{}c@{}}coronavirus\_virus\_\\ vaccine\_pandemic\end{tabular}}        & \cellcolor[HTML]{FFFFFF}\begin{tabular}[c]{@{}c@{}}mental\_depression\_\\ stress\_suicide\end{tabular}          \\ \hline
religion                  & \multicolumn{1}{c|}{\cellcolor[HTML]{FFFFFF}\begin{tabular}[c]{@{}c@{}}francis\_pope\_\\ church\_catholic\end{tabular}}             & \cellcolor[HTML]{FFFFFF}jews\_jewish\_chauvin\_verdict                                                         \\ \hline
technology                & \multicolumn{1}{c|}{\cellcolor[HTML]{FFFFFF}\begin{tabular}[c]{@{}c@{}}tech\_ai\_\\ technology\_watson\end{tabular}}                & \cellcolor[HTML]{FFFFFF}\begin{tabular}[c]{@{}c@{}}bitcoin\_blockchain\_\\ crypto\_cryptocurrency\end{tabular} \\ \hline
art    & \multicolumn{1}{c|}{\cellcolor[HTML]{FFFFFF}\begin{tabular}[c]{@{}c@{}}film\_mckay\_\\ movie\_netflix\end{tabular}}  & \cellcolor[HTML]{FFFFFF}\begin{tabular}[c]{@{}c@{}}museum\_futter\\ \_museums\_mercer\end{tabular}             \\ \hline
economy     & \multicolumn{1}{c|}{\cellcolor[HTML]{FFFFFF}\begin{tabular}[c]{@{}c@{}}companies\_corporate\_\\ business\_investors\end{tabular}}   & \cellcolor[HTML]{FFFFFF}\begin{tabular}[c]{@{}c@{}}fashion\_brands\_\\ clothes\_clothing\end{tabular}           \\ \hline
education                 & \multicolumn{1}{c|}{\cellcolor[HTML]{FFFFFF}\begin{tabular}[c]{@{}c@{}}candidates\_questions\_\\ question\_twentyone\end{tabular}} & \cellcolor[HTML]{FFFFFF}\begin{tabular}[c]{@{}c@{}}education\_students\_\\ teachers\_standards\end{tabular}    \\ \hline
\end{tabular} 
\end{table}

Following this manual classification, we performed a correlation test with the topics around various geoengineering approaches. These results are illustrated in Table \ref{tab:Tab3}, which shows a topic-topic correlation at 95\% CI. For example, `afforestation' and `disaster' related topics have an R$^{2}=0.10$. Similarly, Direct Air Capture (DAC) has correlations with topics in nature (R$^{2}=0.14$) and politics (R$^{2}=0.13$), and so on.

Therefore, supporting our \textit{hypothesis 2 (H2)} that not all topics related to environmental issues equally prompt public interest in geoengineering. Specifically, techno-centric geoengineering options like DAC, MCB, SG, and SAI are more likely to generate interest in news topics on nature, politics, economics, and medical (public health) (see Table \ref{tab:Tab3}).

\begin{table}[htbp]
    \centering
\caption{Correlation between environment-related topics in the news corpus and various geoengineering approaches. We only show values at 95\% significance levels.}
\label{tab:Tab3}
    \begin{tabular}{|>{\centering\arraybackslash}p{0.15\linewidth}|c|c|c|l|l|l|l|} \hline 
 Geoengineering approaches& \multicolumn{7}{|c|}{Manually encoded topics after BERTtopic extraction}\\ \hline 
         &  Nature&  Politics&  Disaster&Energy &Technology &Economy &Medical \\ \hline 
         Afforestation&  &  &  0.10&& & & \\ \hline 
         Biochar&  &  &  & &0.11 & & \\ \hline 
         Direct Air Capture (DAC)&  0.14&  0.13&  & & & & \\ \hline 
         Marine Cloud Brightening (MCB)&  0.12&  &  & & & & \\ \hline 
         Solar Geoengineering (SG)&  &  &  & 0.16& & & \\ \hline 
         Stratospheric Aerosol Injection (SAI)&  &  &  & & &0.11 &0.10 \\ \hline
    \end{tabular}

\end{table}

\subsection{Sentiment dynamics between environmental topics and geoengineering}

\begin{figure*}[htbp]
\centering
\includegraphics[width=0.7\linewidth]{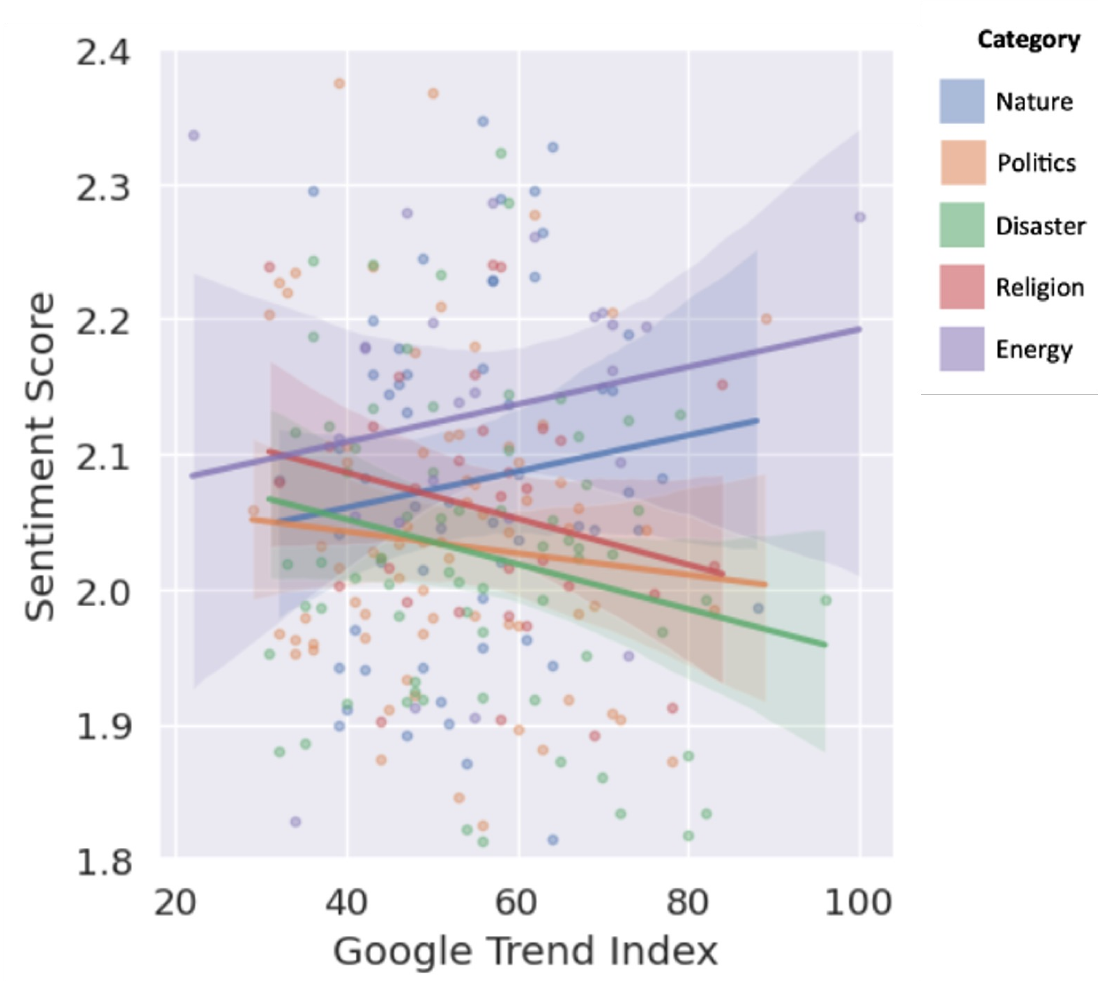}
\caption{\textbf{Relationship between sentiment scores of environment and climate-related news articles and public attention towards geoengineering.} The x-axis shows the Google Trends Index scores, representing public interests in geoengineering, and the y-axis represents the sentiment scores across the five topic categories from the BBC and the NYTimes corpus. The trend line shows the Pearson correlations, with shaded regions showing uncertainties at 95\% levels. The correlation coefficient value for ``Nature" is 0.10, -0.08 for ``Politics", -0.18 for ``Disaster", -0.24 for ``Religion", and 0.18 for ``Energy".} \label{fig:3}
\end{figure*}

Figure \ref{fig:3} shows the relationship between public interests in geoengineering topics (proxied through Google Trends Index) and the sentiment scores of the environment and climate news corpus (classified into 5 manual categories) from the BBC and the NYTimes. The statistical distribution of the computed sentiment scores for our corpus is illustrated in SI figure 4 and SI figure 5. Our results show that categories like nature (R$^2=0.10$ at 95\% CI) and energy (R$^2=0.18$ at 95\% CI) are associated with positive sentiment scores and show an upward trend with increased geoengineering attention (see Figure \ref{fig:3}). 

For topics associated with religion (R$^2=-0.24$) and politics (R$^2=-0.08$), the results show a negative trend at 95\% CI. This can be due to the fact that topics associated with religion and politics have a more complex sentiment-attention relationship with geoengineering approaches that are often shaped by diverse beliefs and values (as reported in \cite{Clingerman_OBrien_2014a, Sovacool_Baum_Low_2022, Corner_Pidgeon_2014, Corner_Pidgeon_Parkhill_2012}).

\begin{figure*}[htbp]
\centering
\includegraphics[width=0.7\linewidth]{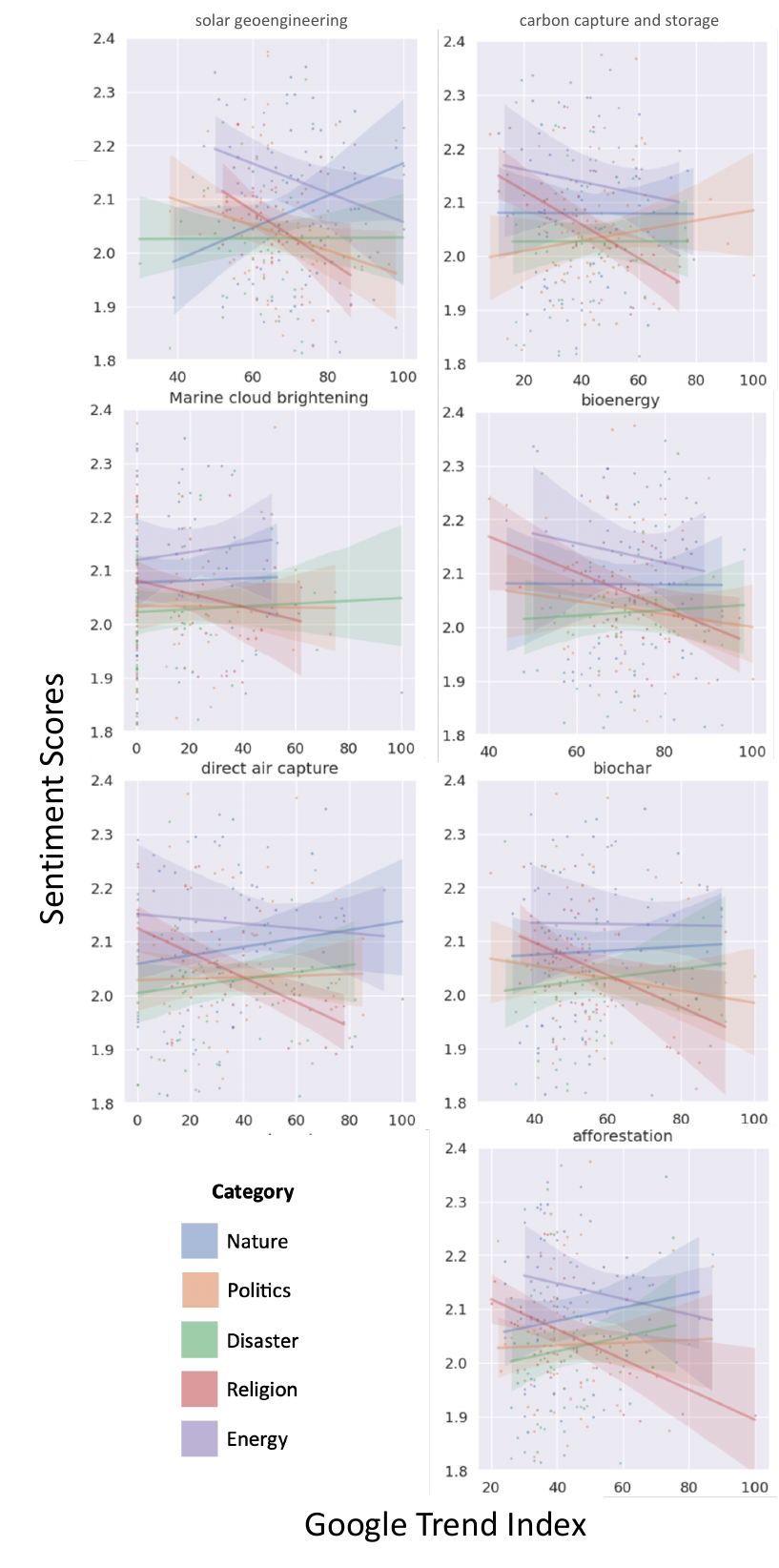}
\caption{\textbf{Sentiment scores of climate-related topics in the news articles and public attention towards the various geoengineering approaches.} The x-axis shows the Google Trend Index, representing interests in types of geoengineering, and the y-axis represents sentiment scores across the 5 categories of climate-related topics in the news data. The trend lines show the Pearson coefficient, with shaded regions showing uncertainties at 95\% levels.} \label{fig:4}
\end{figure*}

\begin{table}[htbp]
    \centering
\caption{Sentiment-attention relationship across the geoengineering approaches and news categories. Pearson correlation coefficients at 95\% significance levels are shown.}
\label{tab:tab4}
    \begin{tabular}{|>{\centering\arraybackslash}p{0.3\linewidth}|c|c|c|c|c|}\hline
 Geoengineering approaches& \multicolumn{5}{|c|}{News categories}\\\hline \hline 
         &  Nature&  Politics&  Disaster&  Religion& Energy\\ \hline 
         Afforestation&  0.11&  &  0.12&  -0.45& -0.18\\ \hline 
         Biochar&  &  -0.12&  &  -0.44& \\ \hline 
         Bioenergy&  &  -0.10&  &  -0.39& -0.14\\ \hline 
         Carbon Capture and Storage (CCS)&  &  0.12&  &  -0.53& -0.13\\ \hline 
         Direct Air Capture (DAC)&  0.12&  &  0.11&  -0.55& \\ \hline 
         Marine Cloud Brightening (MCB)&  &  &  &  -0.25& \\ \hline 
         Solar Geoengineering (SG)&  0.22&  -0.12&  &  -0.42& -0.34\\ \hline 
         Stratospheric Aerosol Injection (SAI)&  0.10&  &  &  0.23& 0.28\\ \hline
    \end{tabular}

\end{table}

Each subplot in figure \ref{fig:4} provides a critical window into the specific relationship of public interests towards geoengineering approaches. For instance, bioenergy, solar geoengineering, and carbon capture and storage (CCS) show varied sentiment-attention correlations, underscoring the importance of tailoring communication strategies to the unique aspects of each geoengineering technique. Table \ref{tab:tab4} shows that religion-related topics are correlated with a negative relationship, although with a weak effect size,  towards all of the geoengineering approaches, except SG. However, attention in SAI shows a positive correlation with energy-related news.

Overall, table \ref{tab:tab4} supports the \textit{hypothesis 3 (H3)} that differential emotional tones in media coverage of climate issues show varying levels of public attention and attitudes towards geoengineering. To explore the relationship between these factors, we report the observed interaction effects in the next section.

\subsection{Interactions and geoengineering attention}



\begin{figure*}[htbp]
\centering
\includegraphics[width=1.0\linewidth]{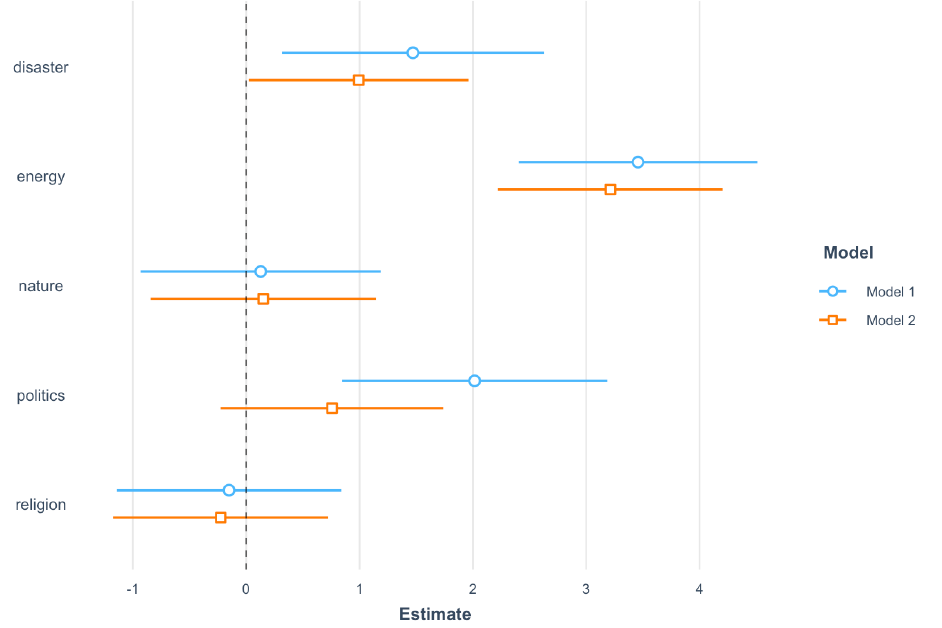}
\caption{\textbf{Estimates of public attention to geoengineering across the five topic categories: disaster, energy, nature, politics, and religion.} Model 1 represents the associations between public interests across the topics (disaster and energy at 99\% CI and politics at 95\% CI). Model 2 shows the relationship between embedded sentiments across the topics (energy at 99\% CI and disaster at 95\% CI). Error bars are shown at 95\% CI, model summary statistics presented in SI Table 2.} \label{fig:5}
\end{figure*}

Figure \ref{fig:5} shows results for regression models 1 and 2. Model 1 shows the correlation between volume of news coverage and public interests, while Model 2 shows the relationship between the embedded sentiments in news articles on geoengineering and the five topic categories: disaster, energy, nature, politics, and religion. Detailed model summary statistics are illustrated in SI Table 2. 

Findings from Model 1 show strong positive relationships between increased public attention to geoengineering and disaster (95\% CI), energy (99\% CI), and politics (99\% CI)-related topics with large associations (see figure \ref{fig:5} and SI Table 2). Model 2 shows relatively large effect sizes for the energy-related sentiment embedding, followed by disaster-related sentiments (see figure \ref{fig:5}). The results show positive sentiments across energy and disaster-related news is associated with relatively higher public attention to geoengineering, at 99\% and 95\% CI respectively. Other topics associated with religion, nature and politics do not show any significant association (see SI Table 2). 

\subsection{Dynamics of geoengineering attention}

\begin{figure}[htbp]
\centering
\subfloat[Media topics and geoengineering interests]{%
  \includegraphics[clip,width=0.9\columnwidth]{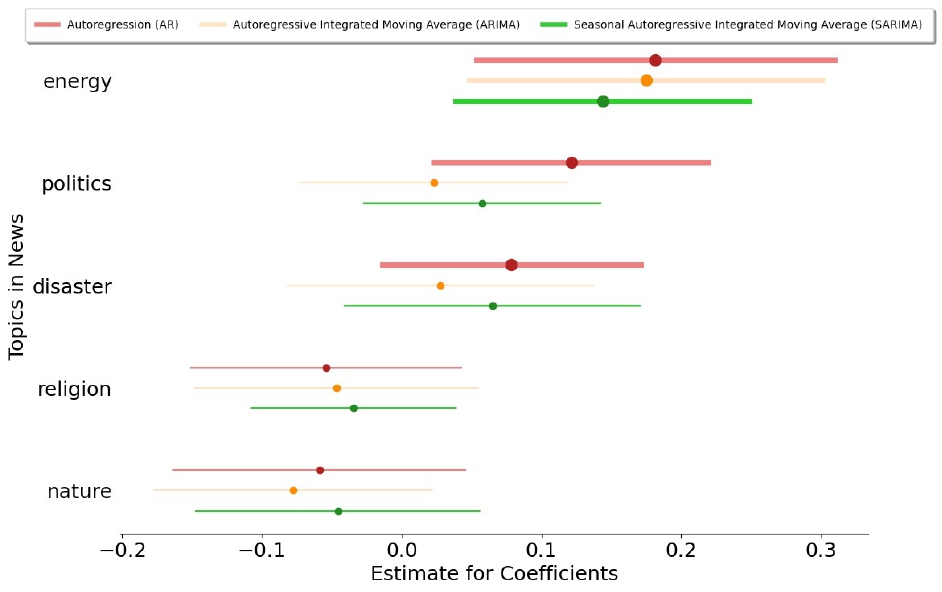}%
}

\subfloat[Embedded sentiments across media topics and geoengineering interests]{%
  \includegraphics[clip,width=0.9\columnwidth]{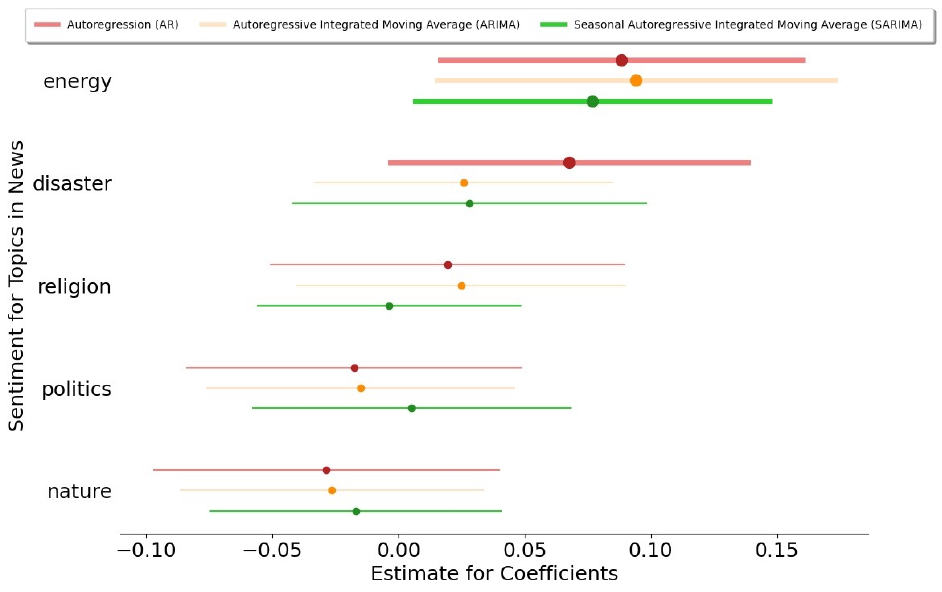}%
}

\caption{\textbf{Associations of various (a) news topic volume and their (b) sentiment scores on public interests in geoengineering, as per the AR, ARIMA, and SARIMA models.} This figure shows the estimated dynamic correlations at 95\% CI with error bars between the BBC and the NYTimes news coverage (across the five topic categories) and their embedded sentiment scores with the Google Trends Index of geoengineering searches. Robustness details for these models are presented in SI Figures 6 - 11.}\label{fig:6}

\end{figure}

Figure \ref{fig:6} shows time-weighted correlations between media coverage of key topics identified in Section 3.4 and public interests in geoengineering, utilizing AR, ARIMA, and SARIMA models. Results show that climate-related news across energy, politics, and disaster topics is positively correlated (at 95\% CI) with an increase in public interest in geoengineering over the lagged observation ($p=4$, see Section 2.3 and SI for model specifications).

Energy and politics-related topics in the news corpus show significant correlations with an increase in geoengineering interests in the AR model with 4 time lags ($p=4$) (see figure \ref{fig:6}). Similarly, the ARIMA model results suggest a more notable and positive correlation between energy news and public awareness compared with other topics. These results illustrate the contemporaneous and lagged correlations ($p=4$, $q=4$) of public interests in geoengineering topics being influenced by energy-related news coverage.

The SARIMA model results show a strong and positive correlation between energy-related news coverage and public awareness of geoengineering. This might be due to the model's ability to account for time-based variations in the data; for example, certain topics associated with climate change (like disasters) are more likely to be covered in specific months when the impacts are felt (see figure \ref{fig:6}). 

We also find that religion and nature topics show a less pronounced correlation with public attention to geoengineering across all models. This might indicate that such topics are covered in ways that do not specifically engage with geoengineering as an emerging climate action technology or that the breadth of discussion on these topics does not directly align to public interest in geoengineering over time.

Figure \ref{fig:6}b shows that the embedded sentiment in energy-related news has a strong positive correlation across the three dynamic models. Similarly, sentiment embeddings across disaster-related topics show relatively weaker but positively correlated results across AR, ARIMA and SARIMA models. 

Additionally, a weak but positive relationship can be observed with sentiment embedding across politics-related topics in the SARIMA model, suggesting the possibility of dynamic relationships that may be attributed to the cyclical nature of political news. The AR and ARIMA modelling results also show a similar relationship for sentiments associated with religion-related news (see figure \ref{fig:6}b). However, we found a negative correlation between sentiment embedding across nature-related topics and dynamic public attention to geoengineering.

\section{Discussion}

Our findings highlight the critical role of the media in shaping and amplifying public interest in geoengineering over time. The significant correlation over time between climate and environmental issue coverage in the BBC and the NYTimes and the geoengineering interests in Google Trends suggests that the public may not consider these themes as unrelated. This insight is particularly valuable for science communicators and educators aiming to open up discussions around the potential role of geoengineering as an emerging portfolio of climate mitigation technology in public discourse (supporting \cite{Corner_Pidgeon_2014, Cox_Spence_Pidgeon_2020, Debnath_Fitzgerald2023, Forster_Vaughan_Gough2020, Buck_2016, Keith_2021})..

The prominence of disasters and energy-related news content corresponds to a correlated increase over time in user searches about geoengineering on Google, suggesting these topics could be a key to a better engagement around geoengineering (see Sections 3.4 and 3.5). Our results also show that the relationship of politics and religion-related topics on the geoengineering discourse is complex, multifaceted, and warrants a nuanced approach. This poses a strategic opportunity to steer public interest in and/or towards geoengineering by carefully selecting the thematic focus of climate communication efforts. This finding aligns with previous data-driven and survey-based studies around geoengineering perceptions that show public opinion is shaped by the uncertainties and risks of these potential technologies and how they are communicated \cite{Corner_Pidgeon_2014, Corner_Pidgeon_Parkhill_2012, Cox_Spence_Pidgeon_2020, Debnath_Fitzgerald2023, gloenvcha2023, Sovacool_Baum_Low_2022, Clingerman_OBrien_2014a}.

We show in Figure \ref{fig:4} and Table \ref{tab:tab4} that different geoengineering approaches have a varied sentiment-attention response mix across five topics, which is an important indicator that not all tone of messaging and public communication can offer the desired engagement response. There is a need for science communicators, educators, public figures, and decision-makers to understand the underlying sentiments towards certain geoengineering topics and devise their strategies accordingly. For example, studies have shown that stratospheric aerosol injection and solar geoengineering are particularly vulnerable to disinformation that demands a careful message framing\cite{Debnath_Fitzgerald2023}. Similarly, nature-based solutions are perceived as more desirable and receive a strong positive tone in the social media discourse \cite{gloenvcha2023}.

Also, using time-series regression models (AR, ARIMA, and SARIMA), news about energy and climate and the sentiments that went along with them showed positive correlations with increased public attention to geoengineering (see figure \ref{fig:6}). Examples of message framing from our data corpus support the above observation, for example a 2019 BBC news article which talks about how renewable energy and technology fixes like geoengineering cannot solely be solutions to climate change \cite{Hornborg_2019}. Similarly, a 2020 New York Times article discusses how, as climate disaster piles up, cooling the planet and reducing fossil fuel emissions through solar geoengineering are attracting new money and attention \cite{Fivella_2020}.

Similarly, politics-related topics show a significant positive relationship with public interests across all models, with AR indicating a strong trend (see figure \ref{fig:6}a). For example, a 2022 New York Times letter to the editor series `Right and Wrong Ways to Address Climate Change' \cite{Hausman_2022} provides a glimpse of the political and public dimensions of GGR. Similarly, the article `Could geoengineering cause a climate war?' in the 2019 BBC Science Focus News discussed conflicting public perceptions regarding geoengineering \cite{Hamilton_2019}.

Disaster-related news also emerges as an important factor in influencing public interests in geoengineering in all models with a relative larger AR coefficient at both news topic and embedded sentiment score levels (see figure \ref{fig:6}). Through observational results, we find that in the context of climate change and extreme weather, disaster and politics are often inseparable \cite{lahsen2022politics}, and our news data corpus also reflects this overlap. For example, the articles `Climate change will bring global tension, a US intelligence report says' in the 2021 BBC News \cite{Corera_2021} and `What if American Democracy Fails the Climate Crisis?' in the 2021 New York Times \cite{Klein_2021} reflect the above empirical observation. Additionally, we observe media-led evidence indicating a growing public interest with increasing climate impacts over time. A 2023 BBC News article reported that search queries related to `climate anxiety' have risen by 150\% as compared to 2022 levels based on the Google Trend Index (see \cite{Gilder_2023}).

We do not see strong associations in our regression results for religion and nature-related topics (see Figure \ref{fig:6}). However, we note a statistically significant negative trend between sentiment scores associated with religion and the Google Trends Index of geoengineering approaches (see Figure \ref{fig:4} and Table 4), indicating that religion is associated with public attention to geoengineering. This supports Clingerman and O'Brien's theoretical arguments \cite{Clingerman_OBrien_2014a}.

\section{Conclusions}

As debates around geoengineering as a potential future climate action strategy intensify, it becomes critical to engage the public and carefully weigh their opinion on the research, development, and possibly deployment of these technologies. However, a lack of public awareness about geoengineering remains a significant barrier, and this study contributed to a better understanding of public attention to emerging geoengineering techniques as a tool for climate action. We used a data-driven and observational approach to evaluate public interests in geoengineering using Google Trends data, as well as how the BBC and New York Times' coverage of climate issues influences public attention using machine learning and dynamic regression analysis.

The results show that energy-related news is more likely to influence public interests with a positive tone in geoengineering over time than disaster- and politics-related news in broader climate communication contexts. We also find religion-related topics to have a negative correlation with interest in geoengineering, though the effect is very small. We also find religion-related topics to have a negative correlation with interest in geoengineering, with a very small effect. Such findings suggest the need for different engagement strategies based on the topic of interest, supporting \cite{Clingerman_OBrien_2014a}. It also sheds light on potential counterstrategies for geoengineering misinformation and conspiracy theories by designing contextualized and interest topic-specific inoculation and message frames. Clearing the (mis)perceptions associated with geoengineering through better communication, message framing, and awareness campaigns can further open up the space for its research and development with greater public participation.

Our observational study has certain limitations, such as the assumption that the Google Trends Index can be a proxy for quantifying public interest in a topic. While this assumption is consistent across current literature, there are questions as to the degree to which it is representative of interests of wider society \cite{Jun_Yoo_Choi_2018, MavraganiOchoaTsagarakis2018}. Similarly, we restricted our data collection strategy to the publicly available big data news repositories of the BBC and the New York Times, thereby narrowing the exploratory scope of the study and potentially introducing certain representation biases due to the possible socio-political leanings of these news agencies. Nonetheless, their combined readership of more than 1.5 billion people globally makes them a helpful case study candidate. In the methodological framework, we use pre-tuned BERTopic and BERT sentiment models, as fine tuning can be resource-intensive. Fine-tuning our custom large language models for climate opinion dynamics tasks remains the topic of ongoing work.

Because of its socio-political influences and high-risk profiles, geoengineering is sensitive to framing effects. While we provide insights using exploratory and data-driven approaches in deciphering public interest in the topic, there is still a need for social science surveys and public engagement activities to improve public awareness, facilitate better understanding, and address framing effects such as partisanship or political ideology in deciding climate mitigation options. \cite{Cox_Spence_Pidgeon_2020, Sovacool_Baum_Low_2022}. We are currently conducting social surveys to measure such framing effects.

This paper presents a step towards a deeper understanding of public interest and perception towards emerging technologies for solar radiation management and greenhouse gas removal, within the broader context of geoengineering. Overall, we contribute to the growing efforts to systematically collect and analyze unstructured data pertaining to public opinion about climate change interventions and the factors that influence them.

\section*{Acknowledgement}
RD acknowledges support from Cambridge Humanities Research Grant, Keynes Fund [JHVH] and Bill and Melinda Gates Foundation [OPP1144]. PZ thanks Downing College and Centre for Climate Repair, University of Cambridge for the summer work placement support. 

\section*{Data availability statement}
The BBC and the New York Times data was extracted using their APIs available in the respective public Github repositories, \url{https://github.com/Sayad-Uddin-Tahsin/BBC-News-API}, and \url{https://github.com/mkearney/nytimes}. Google Trend data is downloaded from publicly available webpage: \url{https://trends.google.com/trends/}. Reproducible codes can be found here: \url{https://github.com/ypzpy/Geoengineering-News}.

\section*{Ethical declarations}
IRB rules do not apply, as no human participant or primary data was used in this research.

\section*{Author contribution}
R.D. and S.D.F conceptualized the project. R.D., P.Z., T.Q., R.M.A., and S.D.F. designed the methodological framework. P.Z. and T.Q. analyzed the data with support from R.D. R.D., P.Z. and T.Q. wrote the original draft. All authors reviewed and edited the manuscript. R.D. provided the overall supervision for the project. R.D. and S.D.F acquired project funding.

\printbibliography



\section*{Supplementary material}

# Supplementary Information (SI)

# Deciphering public attention to geoengineering and climate issues using machine learning and dynamic analysis

Ramit Debnath[1,2], Pengyu Zhang[1], Tianzhu Qin[1], R. Michael Alvarez[2], Shaun D. Fitzgerald[1]

[1]University of Cambridge, Cambridge, CB21PX, UK
[2]California Institute of Technology, Pasadena, 91125, US

**Word2Vec model specifications:**
The distance $D$ between two words $w_1$ and $w_2$ is given by the formula:

$$D(\vec{v}_{w_1}, \vec{v}_{w_2}) = \sqrt{\sum_{i=1}^{n}(v_{w1_i} - v_{w2_i})^2}$$

where $v_{w1i}$ and $v_{w2i}$ are the components of the vectors $v_{w1}$ and $v_{w2}$, and $n$ is the dimensionality of the vector space.

**Regression models (Model 1, Model 2):**

The regression models used in the paper have the following specifications:

$$Geoengineering = \alpha + \sum_{i=1}^{n} \beta_i Topic_i + \epsilon$$

where,
*Geoengineering*: Public interests and attention of geoengineering
*Topic$_i$*: The $i_{th}$ topic score or sentiment score
$n$: Number of the topics
$\alpha$: Intercept for the model
$\beta_i$: Coefficient for $X_i$
$\epsilon$: Error term for the model

**AR, ARIMA and SARIMA models:**

The autoregression (AR) model is a cornerstone of time series forecasting, relying on the premise that past values have predictive power over future values. It is a statistical approach that models the variable of interest using a linear combination of its own previous values. In

the AR model, the current value of the time series is expressed as a function of its past values, with the inclusion of a stochastic error term. The order p of the AR model denotes the number of lagged observations in the model. We set p as 4, and the model formula is shown in the following equation.

$$\begin{aligned}Geoengineering_t =& \phi_1 Geoengineering_{t-1} + \phi_2 Geoengineering_{t-2} \\ &+ \phi_3 Geoengineering_{t-3} + \phi_4 Geoengineering_{t-4} \\ &+ \sum_{i=1}^{n} \beta_i Topic_i \\ &+ \epsilon \end{aligned}$$

The ARIMA (p,d,q) model consists of the AR model in which the current value of the series can be explained as a function of p past values, the integrated model of order d, and the moving average (MA) model in which the current value of the series was explained as a function of current and the q most recent past white noise. We set p as 4, d as 0 (stationary), and q as 4, which leads to the model formula below.

$$\begin{aligned}Geoengineering_t =& \phi_1 Geoengineering_{t-1} + \phi_2 Geoengineering_{t-2} \\ &+ \phi_3 Geoengineering_{t-3} + \phi_4 Geoengineering_{t-4} \\ &+ \theta_1 \epsilon_{t-1} + \theta_2 \epsilon_{t-2} + \theta_3 \epsilon_{t-3} + \theta_4 \epsilon_{t-4} \\ &+ \sum_{i=1}^{n} \beta_i Topic_i \\ &+ \epsilon \end{aligned}$$

We also tailored the SARIMA to time series data with a seasonal pattern that repeats every 52 periods, which in our case corresponds to weekly data in a year. The model is defined by a number of important parameters that show the order of the autoregressive (AR), differencing (I), and moving average (MA) parts for both seasonal and non-seasonal parts.

For the non-seasonal part, our model employs 4 AR terms, 0 differencing, and 4 MA terms. For the seasonal part, it includes 1 seasonal AR term, 0 seasonal differencing, and 1 seasonal MA term. The empirical expression is illustrated in the following equation.

$$\begin{aligned}Geoengineering_t =& \phi_1 Geoengineering_{t-1} + \phi_2 Geoengineering_{t-2} \\ &+ \phi_3 Geoengineering_{t-3} + \phi_4 Geoengineering_{t-4} \\ &+ \theta_1 \epsilon_{t-1} + \theta_2 \epsilon_{t-2} + \theta_3 \epsilon_{t-3} + \theta_4 \epsilon_{t-4} \\ &+ \Phi_1 Geoengineering_{t-52} + \Theta_1 \epsilon_{t-52} \\ &+ \sum_{i=1}^{n} \beta_i Topic_i \\ &+ \epsilon_t \end{aligned}$$

where,

$Geoengineering_t$: Public interests and attention of geoengineering for the time t

$Topic_i$: The $i^{th}$ topic score or sentiment score

$β_i$: Coefficient for $Topic_i$

ϕ: Coefficients for the non-seasonal AR terms ($Geoengineering_t$ term)

θ: Coefficients for the non-seasonal MA terms (error term)

Φ: Coefficients for the seasonal AR term ($Geoengineering_t$ term)

Θ: Coefficients for the seasonal MA term (error term)

n: Number of the topics

$\epsilon_t$: Error term at time t

**SI Table 1.** Descriptive statistics of the news corpus from the BBC and the NYTimes corpus containing geoengineering and climate-related content

| Category | Number of News Articles | | | | | Total | Mean | Median |
|---|---|---|---|---|---|---|---|---|
| | **2018** | **2019** | **2020** | **2021** | **2022** | | | |
| **BBC** | 664 | 1013 | 893 | 1402 | 1361 | 5333 | 1067 | 1013 |
| **NYTimes** | 4508 | 5380 | 4888 | 5604 | 5060 | 25440 | 5088 | 5060 |
| **Total** | 5172 | 6393 | 5781 | 7006 | 6421 | **30773** | 6155 | 6393 |

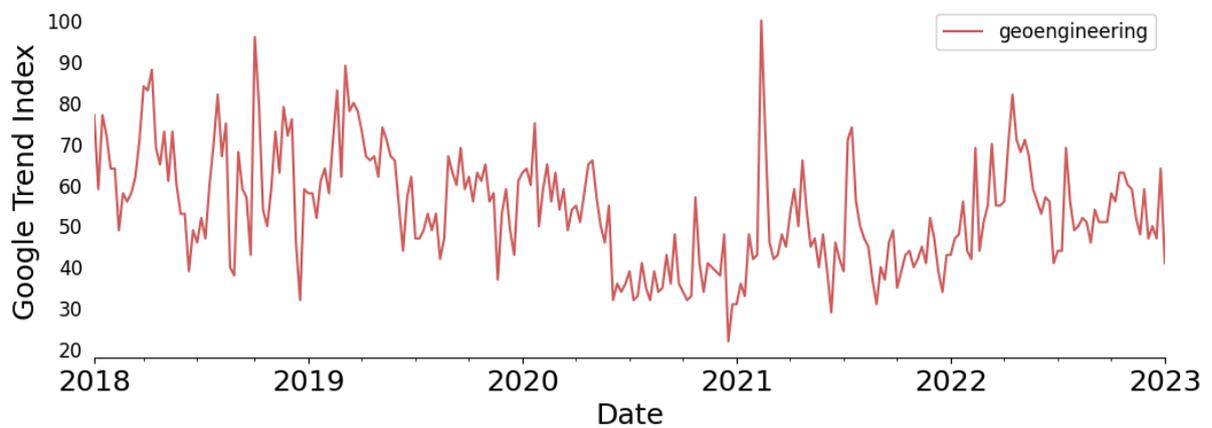

**SI Figure 1:** Google Trend Index timeline for 'geoengineering'. In the paper we used the data between 2018 and 2022.

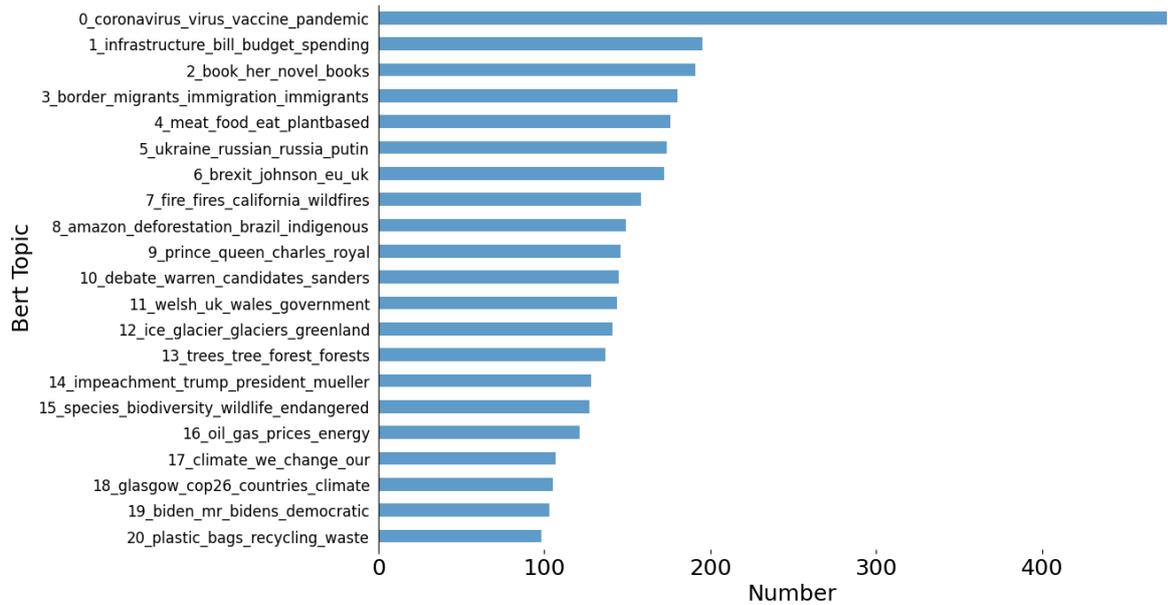

**SI Figure 2:** BERT topic distribution following the topic modelling on the news corpus

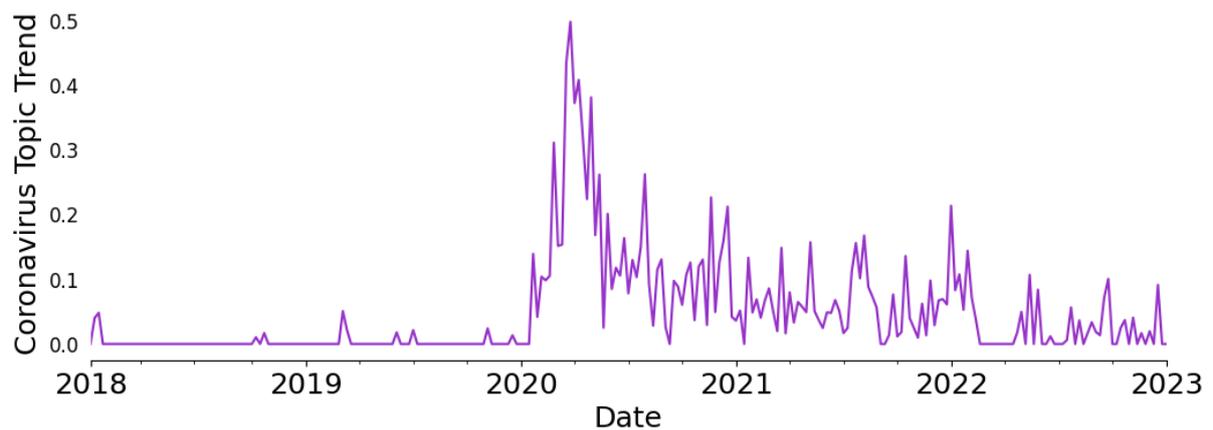

**SI Figure 3:** Example of a BERT topic distribution in the corpus, for the most probable topic, containing COVID-19 related news

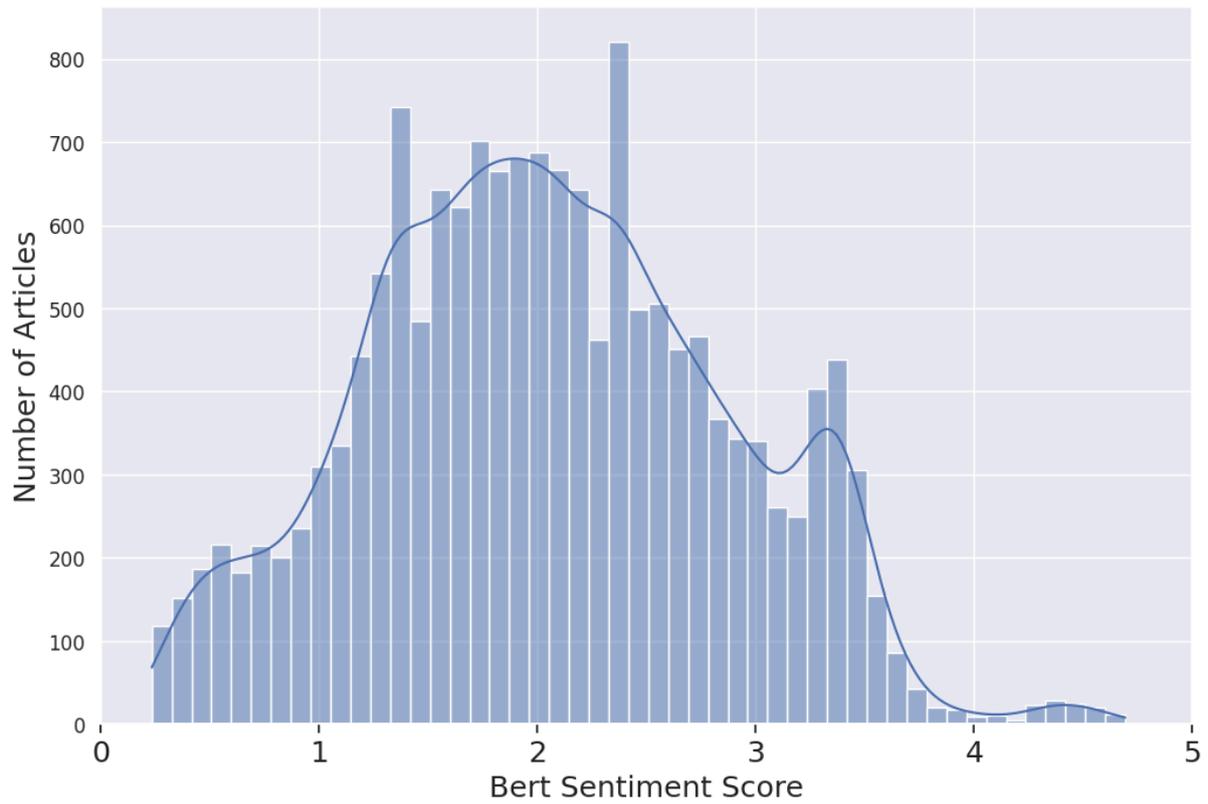

**SI Figure 4:** Cumulative distribution of sentiment scores following BERT sentiment analysis.

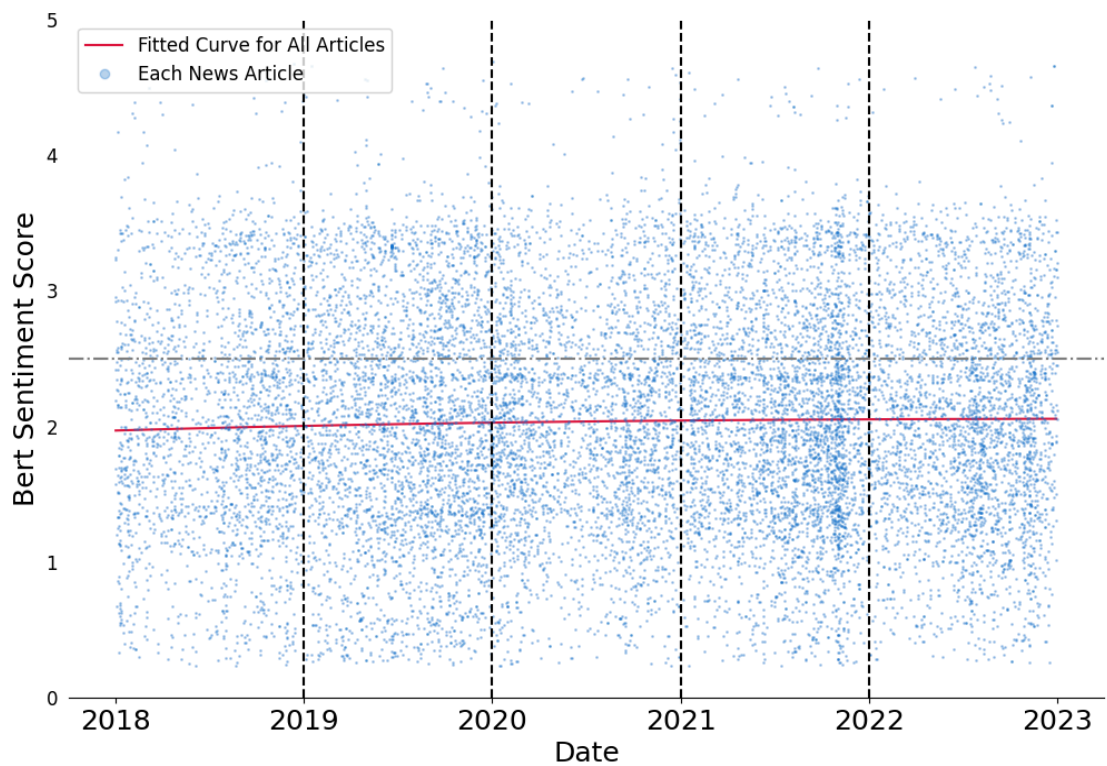

**SI Figure 5:** Spread of the sentiment scores (0 - 5) computed for the news corpus using BERT sentiment.

**SI Table 2.** Regression model results illustrating geoengineering attention correlates at 99% (***), 95% (**) and 90% (*) significance levels.

|  | **Model1** |  | **Model2** |
|---|---|---|---|
| Disaster | 536.15** | Sentiment of Disaster | 1.99** |
| Energy | 1569.42*** | Sentiment of Energy | 7.01*** |
| Nature | 43.49 | Sentiment of Nature | 0.30 |
| Politics | 1123.21*** | Sentiment of Politics | 1.71 |
| Religion | -17.61 | Sentiment of Religion | -1.23 |
| *Intercept* | *27.12*** | *Intercept* | *33.19*** |
| *R2* | *0.15* | *R2* | *0.20* |
| *Adjusted-R2* | *0.14* | *Adjusted-R2* | *0.18* |
| *No. of Observations* | *261* | *No. of Observations* | *261* |

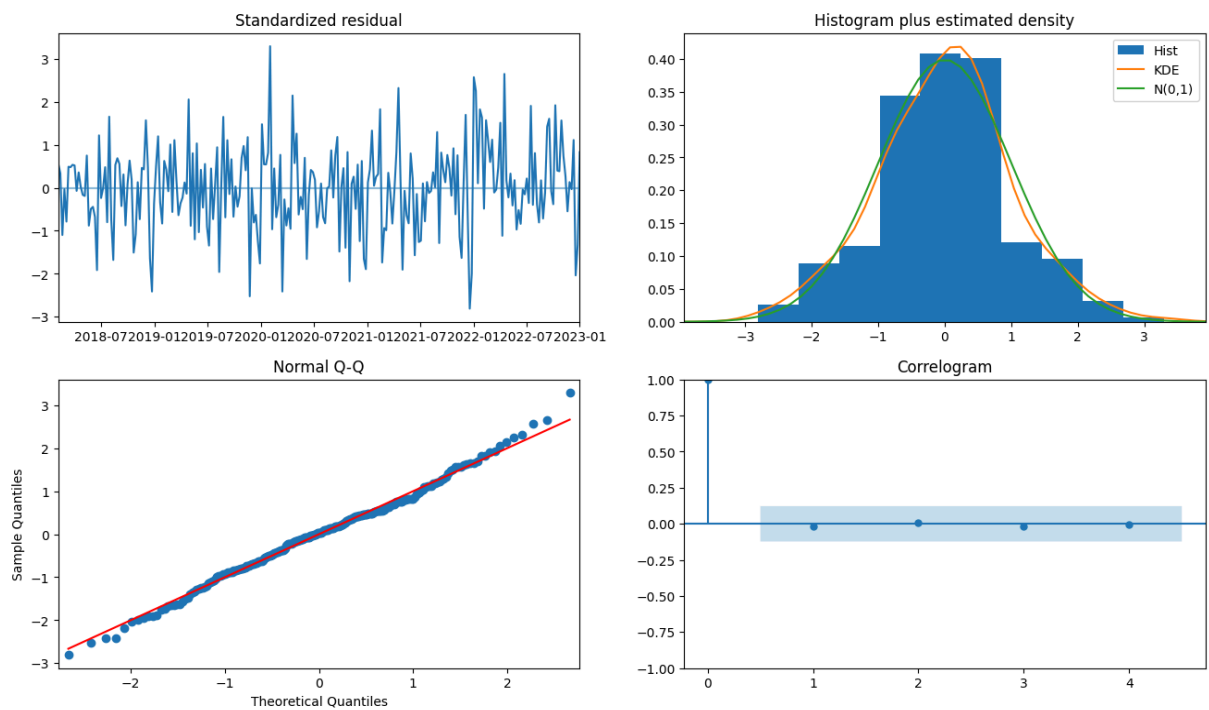

**SI Figure 6:** Autoregression (AR) for topic - 1) Standardized residuals over time; 2) Histogram plus estimated density of standardized residuals, along with a Normal(0,1) density plotted for reference; 3) Normal Q-Q plot, with Normal reference line; 4) Correlogram

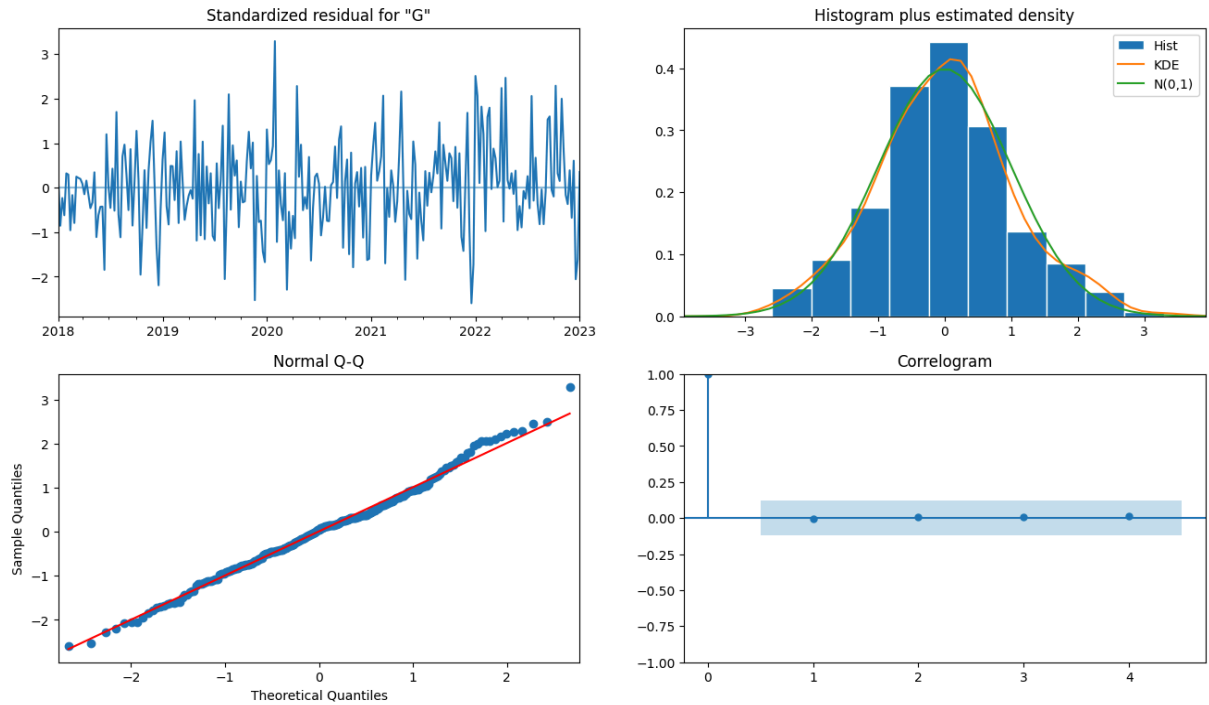

**SI Figure 7:** Autoregressive integrated moving average (ARIMA) for topic - 1) Standardized residuals over time; 2) Histogram plus estimated density of standardized residuals, along with a Normal(0,1) density plotted for reference; 3) Normal Q-Q plot, with Normal reference line; 4) Correlogram

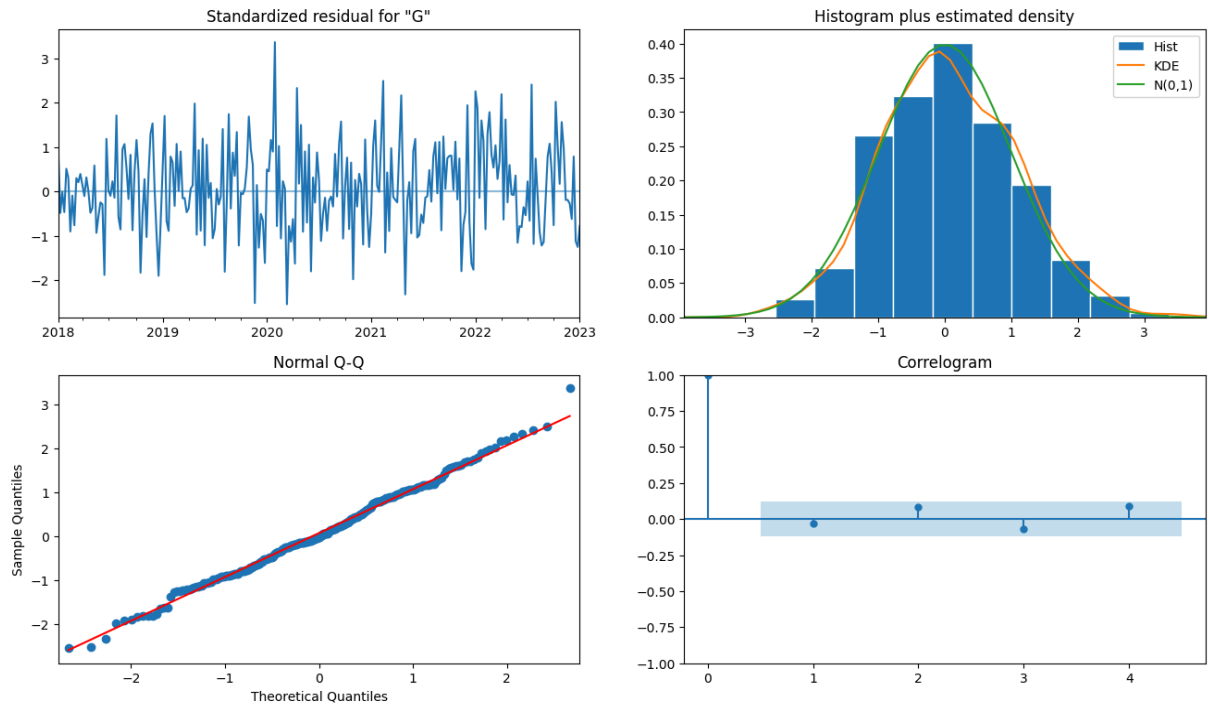

**SI Figure 8:** Seasonal autoregressive integrated moving average (SARIMA) for topic - 1) Standardized residuals over time; 2) Histogram plus estimated density of standardized residuals, along with a Normal(0,1) density plotted for reference; 3) Normal Q-Q plot, with Normal reference line; 4) Correlogram

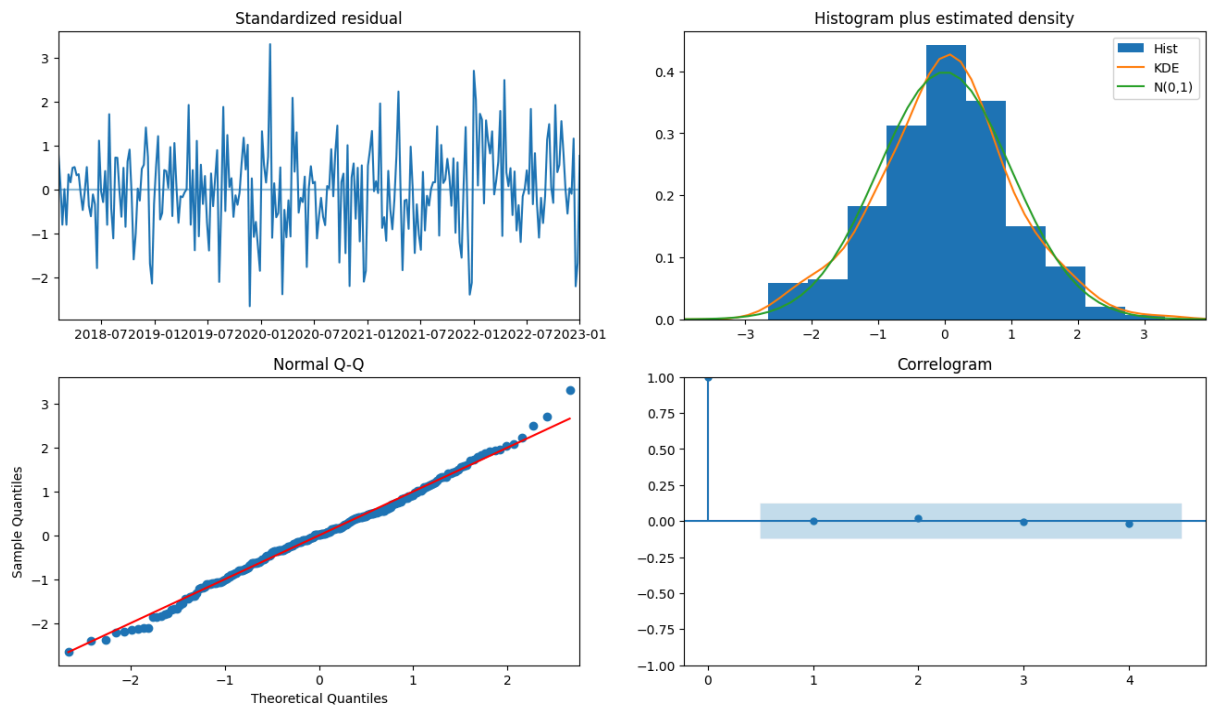

**SI Figure 9:** Autoregression (AR) for topic sentiment - 1) Standardized residuals over time; 2) Histogram plus estimated density of standardized residuals, along with a Normal(0,1) density plotted for reference; 3) Normal Q-Q plot, with Normal reference line; 4) Correlogram

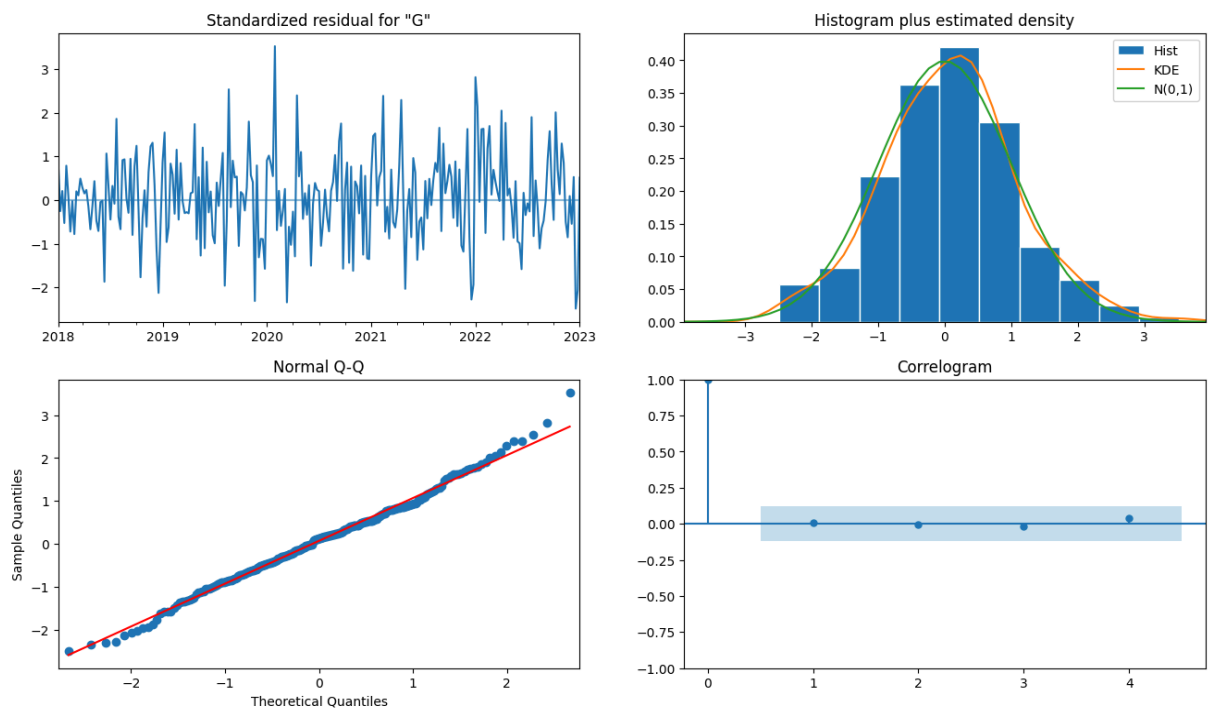

**SI Figure 10:** Autoregressive integrated moving average (ARIMA) for topic sentiment - 1) Standardized residuals over time; 2) Histogram plus estimated density of standardized residuals, along with a Normal(0,1) density plotted for reference; 3) Normal Q-Q plot, with Normal reference line; 4) Correlogram

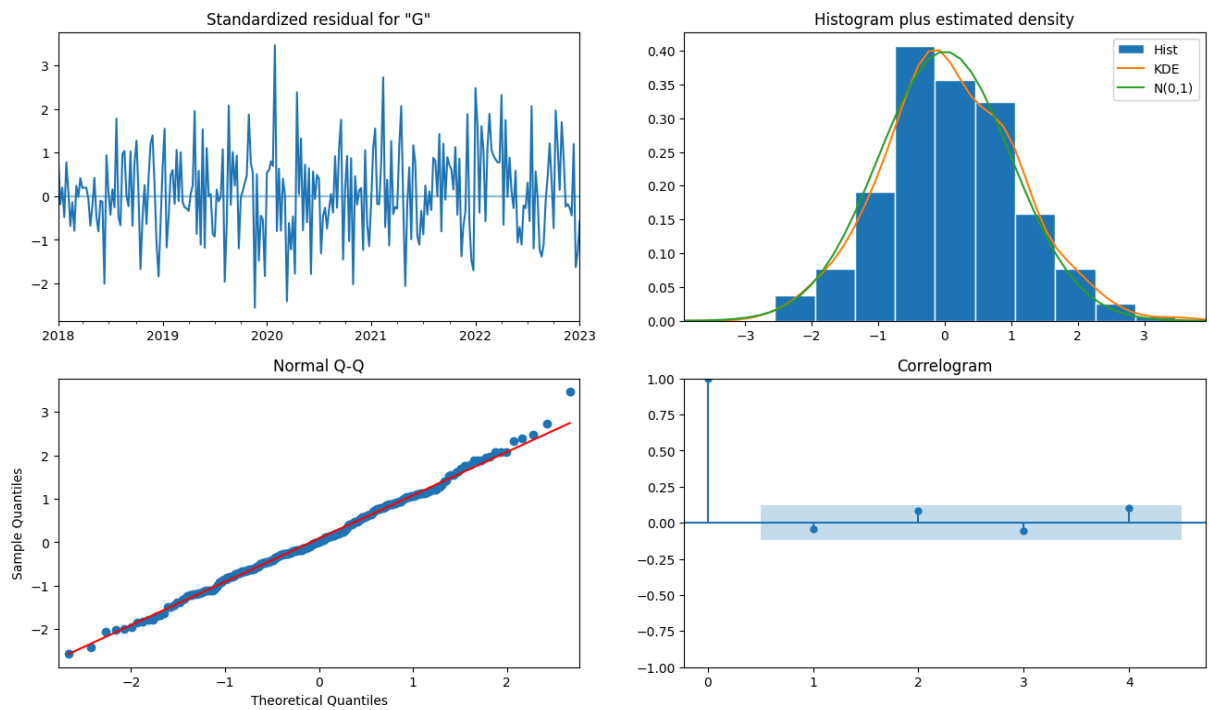

**SI Figure 11:** Seasonal autoregressive integrated moving average (SARIMA) for topic sentiment - 1) Standardized residuals over time; 2) Histogram plus estimated density of standardized residuals, along with a Normal(0,1) density plotted for reference; 3) Normal Q-Q plot, with Normal reference line; 4) Correlogram


\end{document}